# Dynamic Transcript Profiling of *Candida Albicans* Infection in Zebrafish: a Pathogen-Host Interaction Study

Yan Yu Chen[1], Chun-Cheih Chao[2]*, Fu-Chen Liu[2], Po-Chen Hsu[3], Hsueh-Fen Chen[3], Shih-Chi Peng[4], Yung-Jen Chuang[2], Chung-Yu Lan[3], Wen-Ping Hsieh[5], David Shan Hill Wong[1][§]

[1]Department of Chemical Engineering, National Tsing Hua University, Hsinchu 300, Taiwan, R.O.C.
[2]Department of Medical Science & Institute of Bioinformatics and Structural Biology, National Tsing Hua University, Hsinchu 300, Taiwan, R.O.C.
[3]Institute of Molecular and Cellular Biology, National Tsing Hua University, Hsinchu 300, Taiwan, R.O.C.
[4]Nuclear Medicine and Molecular Imaging Center, Chang Gung Memorial Hospital, Taoyuan 333, Taiwan, R.O.C.
[5]Institute of Statistics, National Tsing Hua University, Hsinchu 300, Taiwan, R.O.C.

*These authors contributed equally to this work

[§]Corresponding author

Email addresses:
  YYC: dboring73211@gmail.com
  CCC: u911636@gmail.com
  FCL: ssspeed503@yahoo.com.tw
  PCH: d9580519@oz.nthu.edu.tw
  HFC: g874212@gmail.com
  SCP: escpeng@gmail.com
  DSHW: dshwong@che.nthu.edu.tw
  YJC: yjchuang@life.nthu.edu.tw
  CYL: cylan@life.nthu.edu.tw
  HWP: wphsieh@stat.nthu.edu.tw

# Abstract

*Candida albicans* is responsible for a number of life-threatening infections and causes considerable morbidity and mortality in immunocompromised patients. Previous studies of *C. albicans* pathogenesis have suggested several steps must occur before virulent infection, including early adhesion, invasion, and late tissue damage. However, the mechanism that triggers *C. albicans* transformation from yeast to hyphae form during infection has yet to be fully elucidated. This study used a systems biology approach to investigate *C. albicans* infection in zebrafish. The surviving fish were sampled at different post-infection time points to obtain time-lapsed, genome-wide transcriptomic data from both organisms, which were accompanied with in sync histological analyses. Principal component analysis (PCA) was used to analyze the dynamic gene expression profiles of significant variations in both *C. albicans* and zebrafish. The results categorized *C. albicans* infection into three progressing phases: adhesion, invasion, and damage. Such findings were highly supported by the corresponding histological analysis. Furthermore, the dynamic interspecies transcript profiling revealed that *C. albicans* activated its filamentous formation during invasion and the iron scavenging functions during the damage phases, whereas zebrafish ceased its iron homeostasis function following massive hemorrhage during the later stages of infection. This was followed by massive hemorrhaging toward the end stage of infection. Most of the immune related genes were expressed as the infection progressed from invasion to the damage phase. Such global, inter-species evidence of virulence-immune and iron competition dynamics during *C. albicans* infection could be crucial in understanding control fungal pathogenesis.

# Introduction

    *Candida albicans* is a major fungal human pathogen [1,2]. Although *C. albicans* exists as a commensal organism on the cutaneous and mucosal surfaces of healthy individuals, it is also an opportunistic pathogen that can cause systemic and chronic infections. In immunocompromised patients, *C. albicans* is responsible for a number of life-threatening infections, causing considerable morbidity and mortality. Prior research has identified several *C. albicans* virulence factors, including yeast-to-hyphae morphogenesis, the expression of cell surface adhesins, and extracellular hydrolytic activities [3-5]. However, the mechanism that triggers *C. albicans* to transform from yeast to hyphae has yet to be fully elucidated [6].

    Previous studies of *C. albicans* pathogenesis have suggested several steps that may lead to virulent infection, including early adhesion, invasion, and late tissue damage [7-9]. *C. albicans* may first colonize the host's epithelial surface and persist on expanding, leading to invasion and tissue damage. In the invasion stage, morphogenesis plays a significant role that involves cell transit from ovoid yeast (blastospore) to a filamentous (pseudohyphae and hyphae) form. Finally, the fungus penetrates deep into tissues and organs, causing severe damage and possible host fatality. The underlying molecular events for such a progression are expected to be regulated by the pathogen-host interaction.

    The interaction between *C. albicans* and the host during infection is a complex and dynamic process. *C. albicans* utilizes different strategies to cope with the host's environmental cues for morphological change, proliferation, and escape from the host's immune defenses. The environmental conditions within the host include nutrient and chemical factors, such as the availability of metal ions and amino acids,

reactive oxygen and nitrogen species, host-secreted antimicrobial peptides, antifungal drugs used in therapy, or changes in temperature and pH [10]. Several genome-wide studies have investigated these complex pathogen-host interactions. They have typically utilized a cell line-based system to represent the host environment within an organ or tissue. These approaches include *C. albicans* infections in oral epithelial cells [8], macrophages [11], neutrophils [12], dendritic cells [13], THP-1 human mononuclear cells [14], vascular endothelial cells [15], and human blood [16]. Moreover, there are some studies on *C. albicans* infections in living hosts, e.g. mice [17,18] and rabbits [19]. However, none of these prior studies provided the whole genome response of a living host during *C. albicans* infection.

Recently, the researchers of this study demonstrated that zebrafish (*Danio rerio*) can be a useful model for invasive fungal research. [20]Two-step homogenization/mRNA extraction of a whole zebrafish infected with *C. albicans* could offer a pool of gene transcripts from both the host and the pathogen, enabling the individual estimation of specific gene expression profiles using sequence-targeted probes derived from individual genomes. [21]In the previous study, the surviving zebrafish were sampled at different post-infection time points to obtain time-lapsed zebrafish and *C. albicans* genome microarray data. This was followed by histological analyses. The analysis of the *C. albicans* expression profile data enabled the identification of key regulatory activities at various infection phases. The results indicated that the pathogen genes involved in the extraction of protein-bonded iron, such as ferritin, transferrin, and heme from the zebrafish host, correlated with biological processes that occur in response to infection, the repair of tissue damage, and recovery of lost iron. The progression of *C. albicans* infection in zebrafish may therefore depend on the pathogen's reaction upon sensing a shortage of iron within the host environment. Taken together, our findings suggested that iron competition may regulate *C. albicans* virulence, and that the underlying regulatory scheme may provide the basis to develop potential therapeutic application in the future.

## Results and Discussion

### Dynamic gene expression analysis of *C. albicans* infection in zebrafish

Time-course gene expression analyses in both *C. albicans* and zebrafish elucidated the molecular basis of *C. albicans* infection within a living host. A total of 6,202 genes were found to exist in the expression profile of *C. albicans*. One-way ANOVA detected significant variations across nine time points. Figure S1A illustrates the numbers of significant genes identified at different *p*-values. Using a *p*-value of ≤ 0.01 as the cut-off threshold, the data filter procedures selected a total of 4,827 significant genes for *C. albicans*. As mentioned previously, hyphal morphogenesis is an important virulence factor in *C. albicans*. As a validation to the proposed approach, the present study's findings identified all key genes related to hyphal formation [10], except *NRG1*. Applying a more stringent criterion of a *p*-value less than $0.01/6202=1.61 \times 10^{-6}$ (with Bonferroni correction), the target gene pool was further reduced to total of 1,307 significant genes (Table S1). Gene ontology (GO) analysis of these 1,307 significant genes, based on the *Candida* Genome Database (http://www.candidagenome.org) [23], revealed that the most significant gene functions in terms of their biological process were related to transport, and the most active cellular component is the cell membrane(Table S2). This indicated that membrane transport could represent the key activated function in *C. albicans* as it grows and transforms within zebrafish during infection.

Principal component analysis (PCA) was applied to analyze the dynamic variations in these significant genes using different time points as variables. The scores of the first, second, and third PCs accounted for 79% of the observed variations in the data. The second PC accounted for an additional 12% of the variations, the third PC accounted for 5%, the fourth PC accounted for 2%, and the others accounted for less than 1%. Hence, the first two PCs could account for more than 91% of the variations in the expression dataset.

Figure 1A displays the relative contribution of different time points to the first two PCs. The first PC featured the differential expression between time points 0.5 and 1, and 12, 16 and 18 hours post infection (hpi). The second PC featured the differences between time points 2, 4, 6 and 8 h, and 0.5, 1, 12, 16 and 18 hpi. Histological analysis of the host zebrafish was performed at the exact time points corresponding with the array study. Using the host's liver as a focus, it was found that the pathogen's morphological features correlated well with the three different stages of fungal infection. In Figure 1C, within 1 hpi, *C. albicans* remained predominantly in yeast form and attached to the surface of the zebrafish liver. No tissue invasion was observed in the samples collected at 0.5 and 1 hpi. This period was thus defined as the adhesion phase. *C. albicans* began to form hyphae around 2 hpi, which marked the starting point for the morphological transition *C. albicans*. The fungal invasion into the liver tissue became evident from 4 to 8 hpi; therefore, the time points from 2 to 8 hpi were defined as the invasion phase. Tissue damage and extensive penetration became evident beyond 12 hpi, and the host survival rate decreased rapidly from this time point onward. This result was consistent with the observations reported in the group's previous study [20], and thus, the time points from 12, 16, to 18 hpi were defined as the damage phase. Here PCA analysis and the histological findings were able to classified, in a phenomenological manner, experimental time points into the adhesion, invasion, and damage phases. Nevertheless, the molecular events that occurred in and between stages may be more complex and remained to be investigated.

The scores of the first two PCs accounted for more than 90% of the variations, and Figure S2A displays the scatter plot of the two scores for these significant genes. These genes were then clustered into two major groups (Groups 1 and 2) using the K-mean method and validity index to cluster the principal component scores. Figure 2A shows their expression profiles. There were three subgroups (Subgroups I, II, and III) with different expression patterns in Group 1. Notably, *C. albicans* highly expressed most of the genes in listed Subgroup I in the adhesion phase, but their expression levels decreased when *C. albicans* entered the invasion and damage phases. The genes in Subgroup II had high positive second PC scores, and those in Subgroup III had high negative second PC scores. The expression of the genes in Subgroup II was lowest in the invasion phase, while the expression of the genes in Subgroup III was highest in the invasion phase.

There were also three subgroups (Subgroups IV, V, and VI) in Group 2. In comparison with the aforementioned data, *C. albicans* did not significantly express most of the identified genes in Subgroup IV in the adhesion phase; instead, their expression increased in the invasion and damage phases. The genes in Subgroup V demonstrated the lowest expression levels in the invasion phase, while the genes in Subgroup VI demonstrated the highest expression levels in the invasion phase.

As described previously, *C. albicans* expressed distinct groups of genes during its interactions with zebrafish. Table S4 lists the genes of the key biological processes that demonstrated an association bias toward either Group 1 or Group 2, with a *p-* value of less than 0.1. In particular, the genes related to hyphal morphogenesis, response to chemical stimuli, response to stress, organelle organization, and cell wall organization demonstrated a significant bias toward Group 1. However, the genes with a significant association bias toward Group 2 tended to function in cellular homeostasis and amino acid metabolic processes. Moreover, the response to chemical stimuli, the response to stress, organelle organization and cell walls are very common GO functions highlighted in *C. albicans*. In order to understand the dynamic gene expression patterns associated with hyphal morphogenesis, this study investigated the genes within the hyphae-related GO categories in Group 1. An in-depth review and analysis of the critical genes identified during the analyses are described below.

**Cell surface-related genes of *C. albicans***
The list of significant *Candida* genes contained 135 cell surface-related genes, which were also divided into two groups in the PC subspace. Figure 3A was obtained from the PCA score and load, which reduced the noise from the microarray. This figure showed the genes' expression profiles throughout the time course of the infection. A compiled list of the genes is shown in Table S1.

There were 17 genes that had expression profiles particularly featured in the invasion phase: *COX11, ERG11, orf19.781, ALS3, CNT, orf19.3983, orf19.6435, SMF12, DFG16, orf19.4947, SCO1, PRA1, PLB5, INT1, NGT1* and *PGA32*. Previous studies have demonstrated that *ALS3* and *INT1* encode cell surface proteins related to *C. albicans* adhesion to the host cells [24,25]. *PRA1* encodes a cell surface antigen, and disruption of the *PRA1* gene protects *C. albicans* against leukocyte killing, thus increasing *C. albicans* virulence and organ invasion *in vivo* [26]. *NGT1* encodes an N-acetylglucosamine transporter induced during hyphal growth [27], and *PGA32* encodes a putative GPI-anchored adhesion-like protein that iron transcriptionally regulates [28]. Taken together, these findings indicated the importance of iron uptake and utilization in *C. albicans* virulence.

As shown in Table S4, GO analysis of the biological processes linked to the cell surface-related genes revealed a number of biological processes that were related to hyphal formation and cell wall reorganization, which demonstrated a significant association bias toward Group 1. However, cellular homeostasis was found to demonstrate a significant bias toward Group 2. Seven out of the nine identified genes involved in this process were *RBT5, PGA10, CSA1, SMF12, HMX1* and *FTR1*, all of which are known to be involved in iron homeostasis, and *CRP1*, which has a known function in copper homeostasis. These results suggested that iron transport and scavenging are important activities of *C. albicans* during the infection process.

**Hyphal-related genes of *C. albicans***

The analysis identified hyphal growth as the fungal pathogen's primary process in the invasion and damage phases. There were a total of 122 hyphal-related genes. In agreement with prior research findings [6,8,10] and CGD annotation, most of these genes have been reported to contribute to *C. albicans* virulence in animal models other than zebrafish, and have thus been identified as essential virulent genes for *C. albicans*. These genes could also be divided into two groups in the PC subspace (Table S1). Figure 3B shows their expression profiles throughout the time course of the infection.

Group 1 contained 86 genes, and 31 genes of which, *PLD1*, *INP51*, *SNF2*, *MNT2*, *MNT1*, *MP65*, *NDT80*, *RFG1*, *CEK1*, *CDC24*, *CRK1*, *PHR1*, *CDC42*, *SFL2*, *PMT4*, *CST20*, *KEX2*, *CLA4*, *CHS3*, *SRV2*, *CCR4*, *PMT1*, *SIT4*, *CDC11*, *ALS1*, *CMP1*, *SET3*, *ARP2*, *MKC1*, *HOG1*, and *TOP1*, are manually curated to *C. albicans* pathogenesis. Group 2 had 36 genes, and 12 of which, *IRO1*, *VPS28*, *SOD1*, *TPS2*, *GLN3*, *SSK1*,

*ASH1*, *ACE2*, *TPS1*, *RVS161*, *DFG16*, and *CHK1*, are manually curated to *C. albicans* pathogenesis. There were six genes that had profiles featured particularly in the invasion phase. Among them, it was found that *C. albicans* transiently suppresses *SHA3, FCR1* and *DFG16* in the invasion phase. In terms of gene function, *SHA3* has a known involvement in glucose transport [29], while *FCR1* encodes a putative zinc cluster transcription factor [29]. *DFG16* is required for host tissue invasion [30]. Conversely, *C. albicans* transiently expresses *CDC14, CLB2*, and *BUB2* in the invasion phase. In terms of gene function, *CDC14* is related to cell cycle function, *CLB2* is required for mitotic exit [31], and *BUB2* participates in the mitotic cell cycle spindle orientation checkpoint [31].

Multiple signaling and regulatory pathways control *C. albicans* yeast-to-hyphal transition, which is also related to the biofilm formation. This study identified 122 significant filamentous related genes. Previous studies have mapped many of these genes onto pathways controlling morphological transition. For example, *CST20* and *CEK1* encode components of the mitogen-activated protein kinase (MAPK) signaling pathway [10]. Moreover, *CDC42*, *CDC24*, and *BEM3* gene products function upstream of the MAPK pathway [32] and are expressed during the adhesion phase. *HOG1*, a MAP kinase involved in osmotic, heavy metal, and core stress responses [33,34], is also activated during the adhesion phase. *SSK1* and *CHK1*, which are response regulators of the two-component system, are activated during the damage phase [35,36].

*C. albicans* expressed the downstream target of the protein kinase C (PKC) pathway, *MKC1*, which is involved in contact-induced invasive filamentation, during the adhesion phase. Of the remaining essential biofilm formation genes, *C. albicans* up-regulated *RBT5* and transiently expressed *ALS3* during the invasion phase.

Analyses also identified *CDC11*, *CDC42*, *ACE2*, and *CDC14*. Among then, *C. albicans* expressed *CDC11* and *CDC42* during the adhesion phase, while *ACE2* was expressed during the damage phase. *CDC14* was transiently expressed during the invasion phase.

Of the genes involved in the pH response signaling pathway, *C. albicans* was found to express *RIM21* during the damage phase and *PHR1* during the adhesion phase. Rim21 may function as a sensor of alkaline pH [37], and Phr1 is a putative cell surface glycosidase with involvement in cross-linking of the beta-1,3- and beta-1,6-glucans [38].

The results of this study further revealed the modulation of several components in the *C. albicans* TOR pathway that would occur during the infection of zebrafish. This was significant, as the target of the rapamycin (TOR) signaling pathway regulates a wide variety of cellular processes in response to nutrients. Of these, *C. albicans* expressed *GLN3* during the damage phase, and expressed *RHB1*, *GAP1*, *MEP2* and *SIT4* during the adhesion phase. Significantly, *CDC42, RHO3, RAC1, RHB1, CDC24, BEM3, RDI1, DCK1, LMO1, ACT1, MYO2, MYO5, ARP2, orf19.6705* and *TPM2* were related to actin cortical patch organization, which can potentially regulate the direction of hyphal projection. It was noteworthy that *C. albicans* only expressed *MYO5* during the damage phase; the other genes were expressed during the adhesion phase.

*C. albicans* activated other hypha-related genes, such as *SIN3* and *SET3*, during the adhesion phase. *SIN3* is known to encode a key component of a specific histone deacetylase complex. It can down-regulate true hyphae formation via binding to the *EFG1* promoter [39]. Efg1 is a transcriptional regulator and activates filamentous growth in *C. albicans*. On the other hand, the gene product of *SET3* forms a histone deacetylase complex that is involved in regulating *C. albicans* morphogenesis and virulence [40].

The present study's analyses of *C. albicans* also identified genes related to calcium. It was found that *C. albicans* would express calmodulin (*CMD1*) and the catalytic subunit of calcineurin (*CMP1*). *C. albicans* also up-regulated the expression of *MID1*, which encodes a putative calcium channel of the high affinity calcium uptake system, during the damage phase. *MID1* plays a role in thigmotropism (contact sensing) and the directional hyphae tip response of *C. albicans* [41].

Taken together, the results indicated that *C. albicans* exhibits dynamic gene regulatory controls in many well-known signaling pathways to coordinate various functions during the adhesion phase, but that it will then progressively suppress them. This result was consistent with the finding of yeast-to-hyphal transition occurring in the early stage of infection (Figure 1C).

**Iron-related genes of *C. albicans***

The analysis of cell surface-related genes identified iron transport and scavenging as the primary processes. The expression of these iron-related genes mainly occurred during the invasion and damage phases. Using the literature search approach described in Material and Methods under the specific functional gene list section, the iron-related genes list was found to contain 238 genes, 61 of which had significant expressions. Of these, 25 were in the Group 1 cluster, while 36 were in the Group 2 cluster within the PC subspace (Table S1). Figure 3C provides a list of the genes in these two groups.

*C. albicans* is known to regulate different genes to utilize and uptake iron by sensing the iron concentration in its environment. Accumulated evidence to link these genes' induction to the surrounding iron concentration has been described in previous studies [28,42-49]. Regarding Group 1, previous investigations have described the induction of *KAR2, BGL2, RFG1, SEC12* and *orf19.2452* being observed in a high iron environment, and the induction of *SAP10, GEA2, GCS1, MET3, orf19.1917, orf19.2244* and *orf19.3216* has been observed in a low iron environment. *C. albicans* also progressively suppresses other iron-related genes, including *FRE7* (a ferric reductase gene), *CSR1, MIG1, MAC1, HOG1, SHE3* (a potential regulator for iron-related activities), *RLI1, SOD3* (possibly iron carrying), *HEM15*, and *orf19.5369* (related to heme biosynthesis) during infection. Regarding Group 2, previous studies have reported that the induction of *HMX1, RBT5, FTR1, SOD5, CSA1, FRP1, FTH1, CHA1, GRE2, GCV1, IHD2, SAP99* and *orf19.6530* will occur in a low iron environment, and that the induction of *MDH1, PGA10, CHT2, ISU1, orf19.2048, BIO2, orf19.413, FUM11, SBA1, SPL1* and *HSP60* will occur in a high iron environment. Notably, the present study identified the progressive expression of *VPS28, MYO5* (related to iron uptake), *FRP2, IRO1* (potential iron regulator), *SOD1* and *SOD2* (iron carrying) during *C. albicans* infection. Expression profiles of nine genes were particularly featured in the invasion phase. Among them, it was found that *C. albicans* transiently suppressed *HEM14, SGE1, YHB1, DFG16* and *SEO1*, and that it transiently expressed *ALS3, ERG11, INT1* and *PGA32*, in the invasion phase. Previous investigations have also reported that the induction of *HEM14, YHB1, ERG11* and *PGA32* will occur in a high iron environment, and that the induction of *SGE1, SEO1* and *INT1* will occur in a low iron environment [18,28,42,45-58].

There are at least three known strategies for *C. albicans* to acquire iron from its host: the siderophore uptake system, the reductive system, and the heme-iron uptake system. The primary iron source appears to be the iron-storage protein, ferritin [42]. Many of these genes have been mapped onto these pathways controlling morphological transition. Als3, which is known to play an important role in hyphae formation and invasion, is needed to acquire iron from the host's ferritin. Based on the dynamic gene expression data, it was found that *C. albicans* only transiently expressed *ALS3* in the early invasion phase, while *FTR1* and other ferric reductases (*FRP1*, *FRP2*, and *GRE2*) were progressively induced during the invasion phase. Their expressions were then sustained in the damage phase. These data indicated that the uptake of iron had already started in the early invasion phase. On the other hand, *C. albicans* expressed *HMX1* in the damage phase, and up-regulated *PGA10* and *RBT5* during the invasion and damage phases. This indicated the possible uptake of iron from the host's hemoglobin as the zebrafish suffered massive hemorrhaging during the damage phase. Recently, several studies have reported an iron homeostasis regulatory circuit in *C. albicans*, which includes the transcriptional factors of Sfu1, Sef1, and Hap43 [28,47,59,60]. Although these pathogen genes were not included in our analysis due to the stringent Bonferroni criteria, the dynamic expression profiles obtained in the study were consistent with the reported findings.

**Dynamic gene expression analysis of zebrafish host response during *C. albicans* infection**

The present study applied the same statistical methods to analyze the zebrafish host's dynamic gene expression profile during *C. albicans* infection. The expression profile contained a total of 43,803 probes. Similarly, one-way ANOVA was applied to detect significant gene expression variations across the nine-point time course. Figure S1B illustrates the numbers of significant genes identified at different *p*-values. There were a total of 10,826 significant probes with a *p*-value less than 0.01. Gene annotation was carried out using the Database of Visualization and Integrated Discovery (DAVID) v6.7 [61] and ZFIN [62]. It shall be noted that, unlike the *Candida* pathogen database, there was less bioinformation available for zebrafish. The resulting list of significant genes contained 6,150 genes, including *CXCL12A, CXCL12B, IL12A, INPPL1B, MYCT1, TLR1, TLR18* and *TLR2,* all of which are associated with host defense or immune response mechanisms. The following discussion focused on a list of 576 significant genes (Table S5) that fulfilled the stringent criterion of a *p*-value less than 0.01/43803 (Bonferroni correction).

PCA was applied to analyze the dynamic variations in these significant genes using different time points as variables. The scores of the first PC accounted for 84% of the observed variations in the data, while the second PC explained an additional 9% of the variations. The third PC explained 2%, and the others explained less than 1%. Therefore, the first two PCs could account for more than 93% of the variations in the expression dataset.

Figure 1B displays the contribution plots of these two principal components. The first PC featured the differences in gene expression between time points 0.5, 1, 2 and 4 hpi, and 12, 16 and 18 hpi. The second PC featured the expression differences between time points 4, 6, 8 and 12 hpi, and 0.5, 1, 2, 16 and 18 hpi. This result indicated that the invasion phase in zebrafish was slightly delayed compared to that in *C. albicans*.

Figure S2B illustrates the loci of significant zebrafish genes in the PC subspace. These could be clustered into two main groups according to the validity index. Figure 2B shows the expression profiles of the two groups. The genes in Group 1 (334 probes, 265 genes) demonstrated high expression in the adhesion phase, but this expression started to decline upon entering the invasion and damage phases. The genes in Group 2 (349 probes, 312 genes) demonstrated low expression in the adhesion phase, but this expression increased in the invasion and damage phases.

Gene ontology analysis of these 576 significant zebrafish host genes was performed using DAVID. The analysis revealed that the cluster of genes with the highest enrichment score was related to iron binding (Table S6), which is discussed in a following section. The cluster with the second highest enrichment score was related to hemostasis. The high enrichment score relating to hemostasis was supported by observations of frequent internal bleeding during the experiments. Infection of zebrafish by *C. albicans* may therefore lead to massive hemorrhaging and the expression of genes related to iron metabolism.

**Immune-related genes of zebrafish**
The immune response is a key process for zebrafish to defend themselves against *C. albicans* infection. Of the 576 significant genes identified, 13 had GO related to immune response and immune system development.

Figure 4A shows the expression profiles of these genes throughout the time course of the infection. Group 1 contained three genes; *HBL4, ENPP2* and *VTNB*. Previous studies have demonstrated that *HBL4* and *VTNB* encode components of the extracellular matrix [63-65]. *HBL4* and *VTNB* are involved in damaged cell repair and are significantly expressed during the adhesion phase, but become progressively suppressed in the damage phases. Group 2 contained the other 10 genes, including *LOC796252, FGF21, IRAK3, IL1B, RUNX1, SERPINE1, SLC25A37, SPRY4, TNFA,* and *TNFB*. Among them, *IL1B, TNFA* and *TNFB* are well-known marker genes of the immune response. Further analysis also identified the expression of the growth factor gene *FGF21*, which implied that zebrafish may activate a healing mechanism to repair the damage caused by *C. albicans* infection. This was supported by the observed up-regulation of *RUNX1*, which may promote the production of blood cells following tissue penetration by *C. albicans*. It was noteworthy that the observed immune response occurred immediately after the initial adhesion phase of the infection.

In contrast to previous studies that used cell lines, tissues and organs to pass as a host system for microbial infection, this study utilized a whole organism with complete immune capability. The GO analysis revealed that only a small set of known immune-related genes were significantly expressed the zebrafish. This indicated that the immune response was only one of numerous simultaneous biological activities vigorously regulated in the host during infection. The complex interaction between the immune system and the other physiological processes in the host warrants further investigation in the future.

**Iron-related genes of zebrafish**

The previously mentioned analyses of the *C. albicans* expression profiles identified iron transport and scavenging as important processes during infection. Interestingly, GO analysis of the significant genes in the host zebrafish also indicated that iron homeostasis was an important biological process. Of the 576 significant zebrafish genes, 24 displayed GO associated with iron binding and iron transport.

Figure 4B shows the expression profiles of the genes throughout the infection time course. Group 1 contained 18 genes that the zebrafish progressively suppressed: *CYP51, ZGC:153647, CYP46A1, ZGC:171945, PAH, ZGC:154042, ZGC:55856, CYP4V2, ZGC:56326, SC4MOL, CYP2J30, AMDHD1, CYP2J22, HPX, DPYD, CAT, SC5DL*, and *CYP2J28*. Group 2 contained six genes that were progressively expressed, including: *ZGC:198419, SLC25A37, ZGC:92066, EGLN3, CH25H*, and *PTGS2B*.

It was worth mentioning that the zebrafish progressively suppressed most of the iron-related genes during infection. It should also be noted that most of the iron-related genes were categorized based on informatic annotations instead of experimental data. Nonetheless, one previous wet bench study has identified *SLC25A37* to be related to iron transporter activity and hemooprotein synthesis [66].

**Comparison between *Candida* gene expression profiles in different hosts**

This present study also compared Candida genome expression profiles in other host models to investigate possible virulence genes. There are several published studies providing accessible transcription profiles of *C. albicans* during infection in different living hosts, such as mice and rabbits [17-19,67]. This study chose the *C. albicans* genome-wide data set from Thewes et al. (2007) [18], whose microarray dataset has a complete Candida gene list and a 4 time point (0, 0.5, 3, 5hpi) series profile generated from samples collected from the livers of mice infected with Candida. To ensure consistency, this study applied the proposed method to re-analyze Thewes's gene expression dataset and identify two groups of pathogen genes, including 288 up-regulated genes and 185 down-regulated genes. Figure 5A shows that the total intersection of the conserved genes from this work and Thewes et al. consisted of 130 genes. Figure 5B and Figure 5C show 39 up-regulated genes and 45 down-regulated genes had the same expression profile in both the zebrafish and the mouse models. The entire list is shown in Table S7.

Among the 84 genes with common dynamic expression profiles, the functions of the genes with the top 15 fold change were studied. In this list, *SAP5* is associated with the utilization of protein as nitrogen source and virulence [68]. *RHR2*, regulated by *HOG1* and *NRG1*, is a response to biofilm and hyphal [69]. *SNZ1*, *CHT2*, *ZRT1*, and *PCK1* are associated with hyphal growth [70-72]. *SIM1*, described as a macrophage-down-regulated gene, was repressed both in mice and fish. In terms of the iron-related genes, *SOD5*, *PGA7*, *FTR1*, and *RBT5* were identified. *FTR1* is a ferric reductase, *RBT5* is known as a hemoglobin receptor, and *PGA7* is described as part of a hemoglobin-receptor gene family [73]. *orf19.6084*, *orf19.449*, and *orf19.5848* are described as biofilm-induced genes [74]. There is no description for *orf19.215*, but it is an ortholog to *ScAVO2*, which is related to Tor2p kinase and other proteins.

Interestingly, there were 29 genes out of 1307 genes related to temperature in the fish model, and 14 genes out of 473 genes in mouse model. There were only five temperature-related genes found in both models (*MYO2*, *SRV2*, *PKC1*, *SMT3* and *MKC1*). A close examination of their expression profiles revealed that only *MYO2* was progressively suppressed in both models, and that *C. albicans* up-regulated the other four genes in mice and down-regulated them in fish. This may have been caused by the body temperature difference between mice (36.9 degree Celsius) and zebrafish (28 degree Celsius). In a previous study, *MYO2* was found to be not only related to temperature but also to cytoskeletal polarity, migration, and hyphal growth, therefore *MYO2* may not be directly affected by temperature [75,76].

To summarize, *C. albicans* will turned on a set of the specific virulence genes, which includes genes related to pathogenesis and iron homeostasis, in both fish and mouse models. These genes not only activated the hyphal growth, but also help the pathogen to uptake iron in the host's microenvironment. It should be pointed out that our dataset include the whole body host response to pathogen, but the dataset of Thewes et al. (2007) was an organ-specific response. Hence there must be caution against too much emphasis on the similarity between the datasets. However, these *Candida* genes with common expression profiles in different host species may indeed be truly important during *C. albicans* infection.

## Conclusions

This study provided a novel set of time-lapse genome expression data between the *Candida* pathogen and zebrafish that captured snapshots of dynamic molecular events. It was found that *C. albicans* turned on yeast-to-hyphae morphogenesis to invade the host, and that many cell surface-related and hyphal-related genes were activated during the early stages of infection.

As expected, the zebrafish progressively induced their immune systems to defend against the aggressor. One of the host's immune defenses was to restrict the iron available to the pathogens [77-79]. Many of the iron-related genes were suppressed in the late phase. A contrary example, *SLC25A37*, which is responsible for pumping iron out of the cell for hemooprotein synthesis [66], was consistent with the hypothesis that the zebrafish host was trying to sequester iron at the initial phase of infection, and that its late expression in the damage phase was caused by massive hemorrhaging. Also, the zebrafish expressed immune related genes much more significantly during the late stage of infection. In response to this defense mechanism, the reductive and heme-iron uptake were found to be important strategies for *C. albicans* to acquire iron from the host [42]. The genes responsible were progressively induced, especially after the collapse of the iron restricting mechanism in the fish host. The expressions of *FTR1*, a reductase in the reductive system, and *RBT5*, a hemoglobin receptor in the heme-iron system, were found to be similar in both the fish and mouse models. The microarray profiles were validated using real-time PCR. The expression patterns of the *RBT5*, *CHT2*, *MNT1* and *PMT1 C*. *albicans* genes (Figure 6A) and the *TNFA*, *SLC25A37*, *VTNB* and *CYP4V2* zebrafish genes (Figure 6B) were consistent with those found in the DNA microarray analyses. [22]In summary, this study offered a global view of interspecies interaction between *C. albicans* and zebrafish. It demonstrated the complexity of the crosslinks among the

intra- and interspecies regulatory networks involved in *C. albicans* filamentous formation and the dynamic changes in its iron-acquisition strategies. The comprehensive time-lapsed dataset and findings of this study have enhanced the knowledge on the underlying molecular mechanisms of pathogen-host integrations. Although the interaction between the immune system and other physiological processes of zebrafish require further investigation in the future, this present work may facilitate the identification of core regulatory events to allow the development of novel approaches to attenuate *C. albicans* virulence.

## Materials and Methods

### *Candida albicans* and zebrafish

Adult AB strain zebrafish were used for the experiments. Their maintenance and care after infection were performed according to the procedures described previously [20]. All of the zebrafish use and experimental protocols in this research were reviewed and approved by the Institutional Animal Care and Use Committee of National Tsing Hua University (IRB Approval NO. 09808). *C. albicans* SC5314 strain was used for injection and was prepared as described previously [20].

### Infection and histological analysis

For the microarray analysis, each fish was intraperitoneally injected with $1\times10^8$ *C. albicans* cells suspended in 10 μl sterile 1X phosphate-buffered saline (PBS) after being anesthetized using 0.17 g/ml of tricaine (Sigma). At each time point (0.5, 1, 2, 4, 6, 8, 12, 16 and 18 hpi), three infected fish were collected, sacrificed by immersion in ice water, frozen in liquid nitrogen, and then maintained at –80 °C. Although using whole infected fish to obtain transcriptional profiles could result in the loss of specific information, such as tissue and organ data, it allowed the investigation of the entire genome response during *C. albicans* infection. For the survival assay, the total injected yeast form cell number was $10^8$ cells. For histological analysis, fish were collected at designated time points, and then fixed, sectioned, stained and observed according to the methods used previously [20].

### *C. albicans* and zebrafish RNA purification

The zebrafish infected with *Candida albicans* were treated with Trizol® reagent (Invitrogen, USA), ground in liquid nitrogen using a small mortar and pestle, and then disrupted using a MagNA Lyser system (Roche) that used glass beads (cat. no. G8772-100G, Sigma) shaken at 5,000 rpm for 15 sec. After phase separation was performed by adding chloroform, the total RNA was purified using an RNeasy Mini Kit (Qiagen, Germany). Purified RNA was quantified at OD260nm using an ND-1000 spectrophotometer (Nanodrop Technology, USA) and qualitated using a Bioanalyzer 2100 (Agilent Technology, USA) with an RNA 6000 nano labchip kit (Agilent Technologies, USA).

### Preparation of *C. albicans* custom array and zebrafish array

For the *C. albicans* gene expression analysis, probes were designed using eArray of Agilent technologies. Along the design process, 6,205 *Candida albicans* transcripts (Assembly 21, from the *Candida* genome database,) were uploaded to eArray and designed using base composition methodology. Duplicated target sequences were removed and the resulting 6,202 probes were designed. The custom microarray was manufactured in an 8x15k format by in situ synthesis of the oligonucleotide probes (Agilent Technology). Each array consisted of 6,202 *Candida albicans* specific probes, which were printed in duplicate. The zebrafish gene expression analysis used Agilent zebrafish V2 oligo arrays (Agilent Technology), which were manufactured in a 4x44K format. Control experiments were performed to that ensure the *C. albicans* samples would not cross-hybridize to the zebrafish microarrays and that the zebrafish samples would not cross-hybridize to the *C. albicans* microarrays. In this experiment, the cross-hybridization between different samples and microarrays was not significant (data not shown).

**Microarray experiments**

A total of 1μg of the total RNA was amplified using a Quick-Amp labeling kit (Agilent Technologies, USA) and labeled with Cy3 (CyDye, PerkinElmer, USA) during the *in vitro* transcription process. A total of 0.625μg of Cy3 cRNA for the *C. albicans* array and 1.65 μg of Cy3 cRNA for the zebrafish array was fragmented to an average size of approximately 50 to 100 nucleotides through incubation in a fragmentation buffer at 60°C for 30 minutes. The fragmented labeled cRNA was then hybridized to an oligo microarray at 60°C for 17 h. After washing and drying using a nitrogen gun, the microarrays were scanned using an Agilent microarray scanner (Agilent Technologies, USA) at 535 nm for Cy3. The scanned images were analyzed using Feature extraction 9.5.3 software (Agilent Technologies, USA), and image analysis and normalization software was employed to quantify the signal and background intensities for each feature. The raw data were uploaded onto the NCBI Gene Expression Omnibus Database (GSE32119).

**Microarray data analysis**

The set of zebrafish arrays were processed separately from the *C. albicans* arrays. The Loess normalization curve for the *C. albicans* data was fitted using every probe on the array, while the Loess curve for the zebrafish data was fitted using the 485 control probes. After the pre-processing step with Loess normalization, the mean intensity levels for each of the 6,202 pairs of duplicated probes on the *C. albicans* array were calculated as representatives. For each of the probes, one-way ANOVA was adopted to screen for significant time variations across the nine time points in either set of data. The significance level was adjusted using the Bonferroni correction to achieve a family-wise error rate of 0.01. The adjustment resulted in 1,307 significant probes (out of the total 6,202) for the *C. albicans* data (Table S1) and 683 probes (out of the total 43,803) for the zebrafish data (Table S5).

**Principal component analysis and cluster number**

Principal component analysis (PCA) uses an orthogonal transformation to convert a set of sample observations of potentially correlated variables into a set of values of uncorrelated variables called principal components [80]. In this study, PCA was used to analyze the dynamic variations of the total significant genes by treating different time points as variables and different genes as samples. There were ten PCs that could explain the microarray variations. The principal component scores of each gene could be clustered in the PC subspace using the K-means method based on a validity index [81]. The validity index minimized the intra-distance and maximized the inter-distances of the clusters to determine the number of clusters.

**Specific functional gene list**

To collect all possible receptors and sensors on the *C. albicans* membrane, text-mining in the GCD was performed, using the keywords "plasma membrane", "cell surface", and "cell wall", in order to research the GO terms "transport" and "membrane" from Table S1. The specific keywords identified a total of 533 genes, which were denoted as cell surface-related genes.

Hyphal-related biological processes were also researched by text-mining using the keywords "filamentous growth", "hyphal growth", and "pseudohyphal growth". This identified 431 genes, which were named filamentous related genes.

As shown in Table S4, the significant biological processes demonstrating the bias toward Group 2, progressive expression, was cellular homeostasis. Conducting searches for possible previously identified iron-induced genes, iron phenotypes genes, predicted iron-related genes [18,46,50-55], iron-responsive regulators [56-58], iron uptake genes [42], iron utilization genes [45,47-49] and iron-modulated genes [28] obtained 238 genes, which were named iron-related genes.

**Binomial test for group gene ontology annotation**

To determine whether a process was substantially biased toward a group, the *p*-value was calculated using binomial cumulative distribution, under the assumption that the significant genes of a specific process were randomly distributed between two groups. *q* represented the probability of finding a gene in Group 1. For example, there were 1,307 significant genes in *C. albicans*, of which 734 were in Group 1 and 573 were in Group 2; hence, $q = 734/1307$. *n* represented the total number of genes related to a specific biological process, such as $n=55$ for the DNA metabolic process. *m* represented the number of genes related to this process in Group 1. The *p*-value of bias toward Group 1 was therefore defined as the cumulative probability of *m* or fewer genes being distributed into Group 1, when *n* genes are randomly distributed into Group 1 and Group 2 according to the probabilities *q* and *1-q*:

$$p-value = 1 - \sum_{i=0}^{m} \binom{n}{i} q^i (1-q)^{n-i}$$

**RNA isolation and quantitative real-time PCR**

The control fish and the infected fish (infected with 10 μl 1×10$^8$ CFU/ml *C. albicans* cells) were sacrificed by immersion in ice water at different time points and homogenized in liquid nitrogen. The total RNA was extracted using TRIzol reagent according to the manufacturer's instructions (Invitrogen), and the RNA quality was analyzed by gel electrophoresis. For cDNA synthesis, 6.25 μg of the total RNA was pretreated with DNase I (Invitrogen) and then reverse transcribed using the SuperScript III enzyme (Invitrogen) in a 50 μl reaction mixture. Quantitative real-time PCR was carried out using the 7500 real-time PCR system (Applied Biosystems). The primers used in this study are listed in Table S8. Briefly, each 15 μl reaction mixture contained 25 ng cDNA, 7.5 μl SYBR green PCR master mixture (Applied Biosystems), 0.2 μM forward primer, and 0.2 μM reverse primer. The reactions were performed using 1 cycle of 95°C for 10 min followed by 40 cycles of 95°C for 15 s and 60°C for 1 min.

**Author Contributions**
Chun-Cheih Chao and Fu-Chen Liu provided the zebrafish and conducted histological analyses, and Po-Chen Hsu provided the C. albicans strains. Hsueh-Fen Chen and Fu-Chen Liu performed the real-time PCR assays. Yan-Yu Chen and Shih-Chi Peng were responsible for microarray analysis. Yan-Yu Chen was responsible for statistical calculations, while Yan-Yu Chen and Po-Chen Hsu conducted the gene analyses. David Shan-Hill Wong and Wen-Ping Hsieh supervised the computational and analytical aspects of this study. Yung-Jen Chuang and Chung-Yu Lan designed and supervised the experimental aspects. All authors were involved in the preparation and revision of this manuscript.

## Acknowledgments

This study was supported by grants NSC 100-2627-B-007-001, NSC100-2627-B-007-002 and NSC100-2627-B-007-003 from the National Science Council (Taiwan, Republic of China). The authors would also like to thank the NTHU-NHRI Zebrafish Core Facility for its support.

# Figures

**Figure 1 - PC load bar plot and histology of zebrafish infected by *C. albicans***

(A) The pathogen's load contributions to the first principal component indicated that the differences in expression between 0.5 and 1 and 12, 16 and 18 hpi represented the axes of largest variation. The load contributions to the second principal components featured the gene expression differences between time points 2, 4, 6 and 8 hpi and 0.5, 1, 12, 16 and 18 hpi. Two principal components sufficiently explained more than 93% of the variations of the zebrafish expression profiles. (B) The host's load contributions to the first principal component indicated that the differences in expression between time points 0.5, 1, 2 and 4 hpi and 12, 16 and 18 hpi represented the axes of largest variation. The load contributions to the second principal components featured the gene expression differences between time points 6, 8 and 12 hpi and 0.5, 1, 2, 4, 16 and 18 hpi. (C) The histology of zebrafish liver indicated that *C. albicans* mainly grew in yeast form at 1 hpi and began to attach to the zebrafish liver. 0.5 and 1 hpi were thus defined as the adhesion phase. *C. albicans* began to transit to the hyphal form at 2 hpi. Invasion to the liver was evident at 4 to 8 hpi. Hence, time points 4 to 8 hpi were defined as the invasion phase, with 2 hpi defined as a transition point for morphogenesis. Beyond 12 hpi, tissue damage, more extensive penetration, and fish death occurred. Time points 12, 16 and 18 hpi were defined as the damage phase.

**Figure 2 - Significant genes profile of *C. albicans* and zebrafish**

The PCA result of *C. albicans* could be classified into two main groups. The heat maps (A) revealed that *C. albicans* progressively suppressed most genes in Group 1 and progressively expressed most genes in Group 2. The genes near the top and bottom of the circle were either over-expressed or suppressed in the invasion phase. The PCA result of the zebrafish could be classified into two groups. The heat maps (B) of the two groups revealed that, in Group 1, the zebrafish progressively suppressed most genes and, in Group 2, progressively expressed most genes.

**Figure 3 - Expression profiles of the cell surface-related, hyphal-related, and iron-related genes in *C. albicans***

The list of significant genes contained 135 cell surface-related genes. PCA divided these into two main groups. (A) shows the cell surface-related gene expression profile. 16 genes, *COX11*, *ERG11*, *orf19.781*, *ALS3*, *CNT*, *orf19.3983*, *orf19.6435*, *SMF12*, *DFG16*, *orf19.4947*, *SCO1*, *PRA1*, *PLB5*, *INT1*, *NGT1* and *PGA32* were particularly featured in the invasion phase. The list of significant genes also contained 122 hyphal-related genes. PCA divided these into two main groups. (B) shows the hyphal-related gene expression profile. Six genes were particularly featured in the invasion phase. Of these, *C. albicans* transiently suppressed *SHA3*, *FCR1* and *DFG16* during the invasion phase. Analysis of the cell surface-related genes identified cellular hemostasis as a primary process on the cell surface, and most of these genes were iron homeostasis-related. The list of significant genes contained 61 iron-related genes.

PCA divided these into two main groups. (C) shows the iron-related gene expression profile. Nine genes were particularly featured in the invasion phase. *C. albicans* transiently suppressed *HEM14*, *SGE1*, *YHB1*, *DFG16* and *SEO1*, and transiently expressed *ALS3*, *ERG11*, *INT1* and *PGA32* in the invasion phase.

**Figure 4 - Expression profiles of the zebrafish immune-related and iron-related genes**

The immune response is a key process for zebrafish to defend themselves against *C. albicans* Of the 576 significant genes identified, 13 genes had GO related to the immune response and immune system development. PCA divided these into two main groups. (A) shows the immune-related gene expression profile. Analysis of the *C. albicans* expression profiles identified iron transport and scavenging as important factors during infection. GO analysis also revealed that iron homeostasis was an important biological process among the significant genes. There were 24 genes in the 576 significant zebrafish genes that had GO associated with iron binding and iron transport. PCA divided these into two main groups. (B) shows the iron-related gene expression profile.

**Figure 5 –Comparison of gene expressions of mouse and zebrafish models during *C. albicans* infection**

The present work was compared with the *C. albicans* genome wide data set in Thewes et al. (2007) [18]. (A) shows that the intersection of this work and that of Thewes et al. had 130 genes. The proposed method was used to re-analyze the gene expression profiles and cluster them into two groups that contained 288 up-regulated genes and 185 down-regulated genes. (B) and (C) show that 39 up-regulated genes and 45 down-regulated genes had the same expression profile in the zebrafish model.

**Figure 6 – Validation of expression patterns by real-time polymerase chain reaction**

(A) The *C. albicans* gene expression patterns (*RBT5*, *CHT2*, *MNT1*, and *PMT1*) revealed by the DNA microarray are shown on the top. Real-time PCR of *C. albicans* genes (RNA isolated from three infected zebrafish per group) are shown in the bottom. *C. albicans* expressed *RBT5* to use iron in the invasion and damage phase (after 6 hpi), expressed *CHT2* in the late stage (after 12 hpi), and activated *MNT1* and *PMT1* in the adhesion phase (0.5 and 2 hpi). (B) The zebrafish gene expression patterns (*TNFA*, *SLC25A37*, *VTNB*, and *CYP4V2*) revealed by the DNA microarray are shown on the top. Real-time PCR of the zebrafish genes (RNA isolated from three infected zebrafish per group) are shown on the bottom. The zebrafish progressively expressed *TNFA* and *SLC25A37* during the infection and turned on *VTNB* and *CYP4V2* in the adhesion phase (0.5, 1 hpi). The profiles of the real time PCR were identical and could validate the present genome microarray in both the host and the pathogen. The asterisks above the bar indicate that the level of target gene expression was significantly different from the levels of expression in the other time points: * = $P < 0.2$; ** = $P < 0.05$; *** = $P < 0.01$.

## Supporting Information

**Figure S1 –*C. albicans* and zebrafish genes at different p-values**

(A) Venn diagram of the significant genes in *C. albicans*. There were a total of 4,827 significant genes with a *p*-value of less than 0.01, and there were 4,827 significant genes with a *p*-value of less than 0.1/6202 (Bonferroni correction). This study focused on 1,307 genes that fulfilled the stringent criterion of a *p*-value less than 0.01/6202. This list included several genes related to *C. albicans* hyphal formation, such as *MIG1*,

*RFG1*, *ECE1*, and *CEK1*. (B) Venn diagram of the significant genes in zebrafish. There were a total of 36,292 significant genes with a *p*-value of less than 0.01, and there were 6,150 significant genes with a *p*-value of less than 0.1/43803. This study focused on 683 genes that fulfilled the same criterion used in the *C. albicans* analyses; namely, a *p*-value less than 0.01/43803 (Bonferroni correction). This gene list included key genes related to the immune response, such as *IL1B*, *TNFA*, and *TNFB*.

(TIFF)

**Figure S2 – Clustering of significant genes in the PC subspace**

(A) A scatter plot of the PC scores of the first two principal components demonstrating that the significant *C. albicans* genes could be classified into two main groups and six sub-groups. (B) A scatter plot of PC scores of the first two principal components demonstrating that the significant *C. albicans* genes could be classified into two main groups.

(TIFF)

**Table S1 – *C. albicans* significant genes list**

Applying a stringent criterion of a *p*-value less than $0.01/6202=1.61 \times 10^{-6}$ (with Bonferroni correction) further reduced the target gene pool and identified a total of 1,307 significant genes. The gene list also shows classifications according to groups in the PC subspace and cell-surface, iron and hyphal related functions.

(XLS)

**Table S2 – *C. albicans* gene ontology analysis**

Gene ontology annotation of 1,307 genes performed using the Candida Genome Database and the GO Slim Mapper terms "biological process" and "cellular components". The most significant biology process GO set related to transport. The most significant cellular component GO set related to the cell membrane.

(XLS)

**Table S3 – GO comparison of total *C. albicans* significant genes group**

The table shows the GO of identified biological processes and demonstrating a significant bias in Group 1 or Group 2. Binomial distribution was used to calculate the *p*-value, assuming that significant genes of a specific process were randomly distributed between two groups.

(XLS)

**Table S4– GO comparison of *C. albicans* cell surface related genes group**

The table shows the GO of identified biological processes of *C. albicans* cell surface related genes and demonstrating a significant bias in Group 1 or Group 2.

(XLS)

**Table S6 – Zebrafish gene ontology analysis**

Gene ontology analysis result of 576 significant zebrafish host genes using DAVID.

(XLS)

**Table S7 – Identical expression profile of *C. albicans* genes in mice host**

The proposed method was used to re-analyze the gene expression profiles in the *C. albicans* genome wide data set in Thewes et al. (2007) [18] and cluster them into two groups of 288 up-regulated genes and 185 down-regulated genes. Among them, 39 up-regulated genes and 45 down-regulated genes had the same expression profile in the zebrafish model. The 84 genes are shown in this table.

(XLS)

**Table S8 –Sequences of primers used for real-time quantitative PCR**
The primers used in this study are listed in this table.

(XLS)

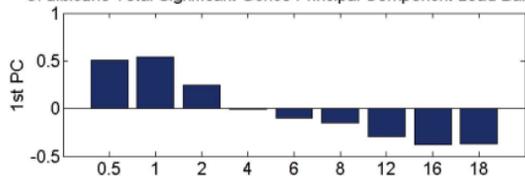
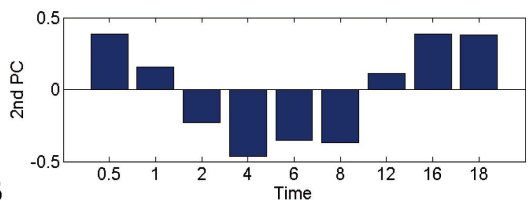
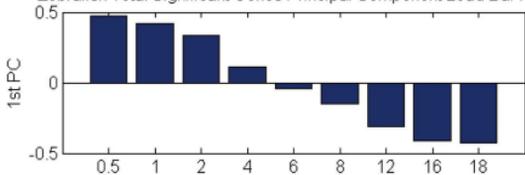
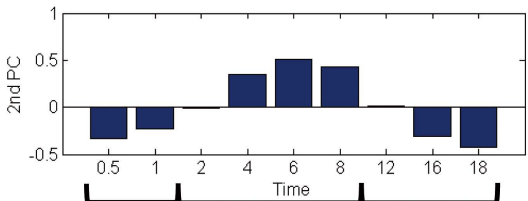
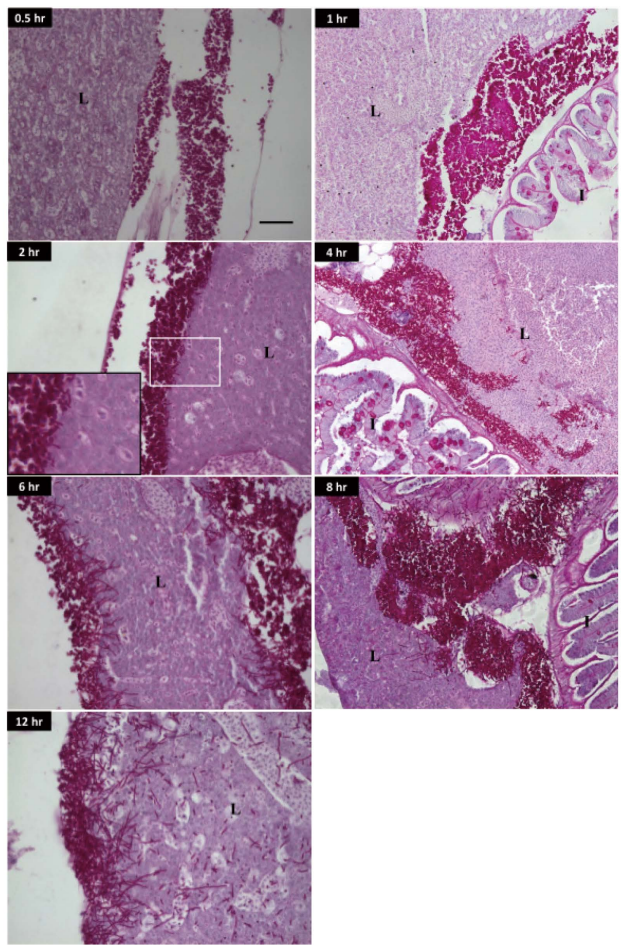

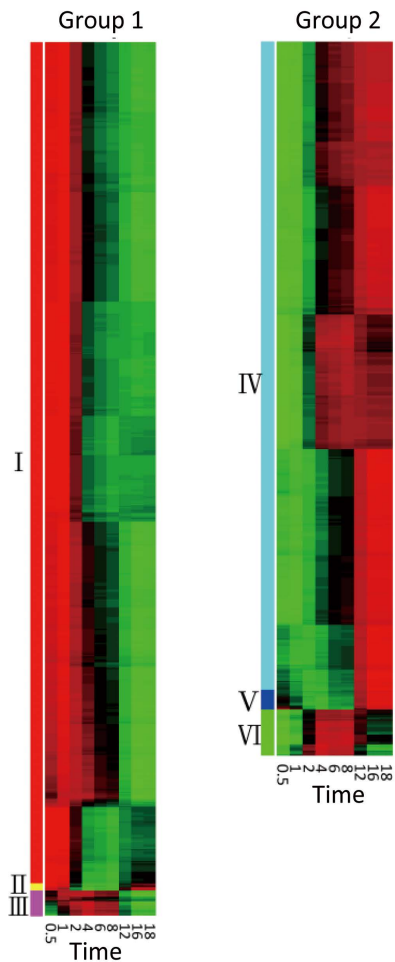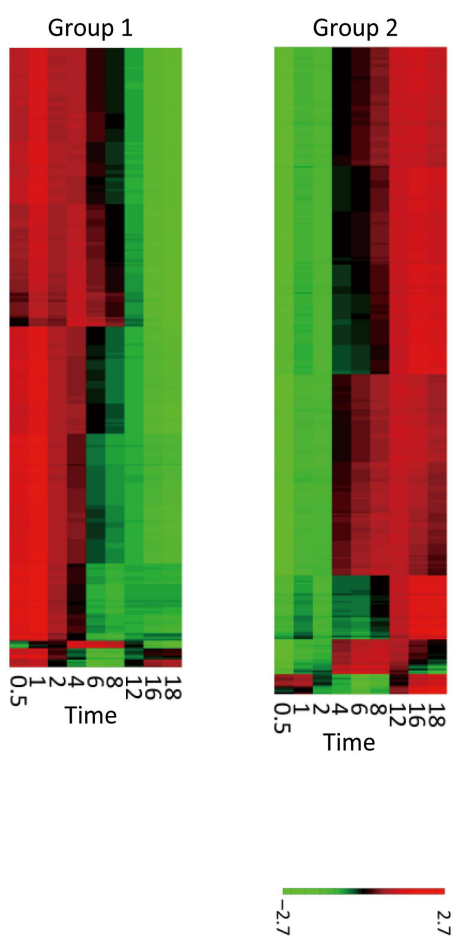

**A** Expression profile of *C. albicans* cell surface related genes

**B** Expression profile of *C. albicans* hyphal related genes

**C** Expression profile of *C. albicans* iron related genes

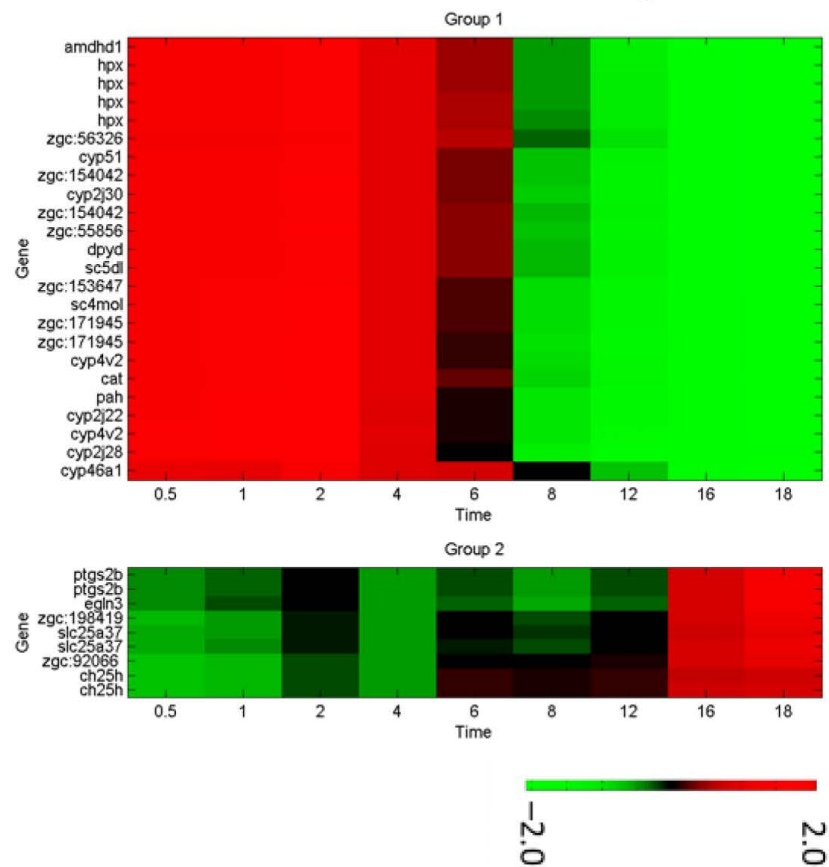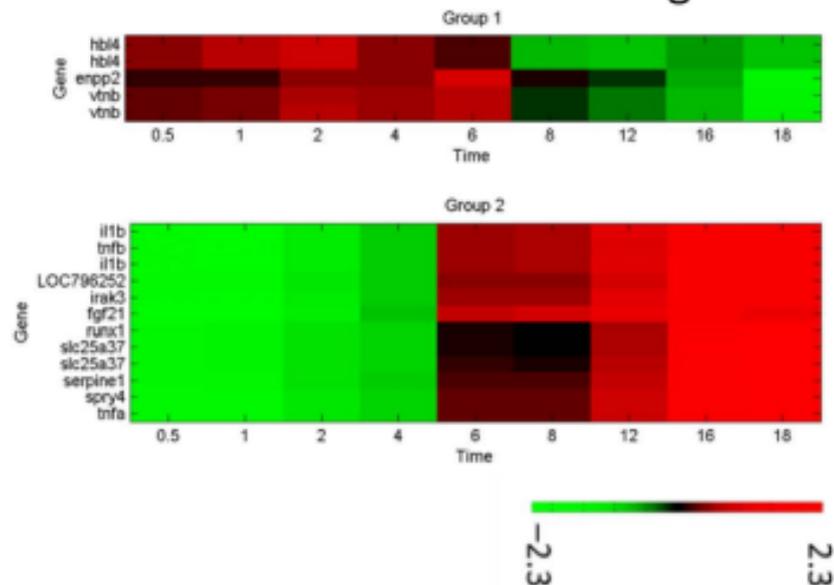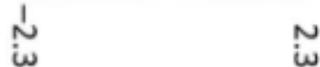

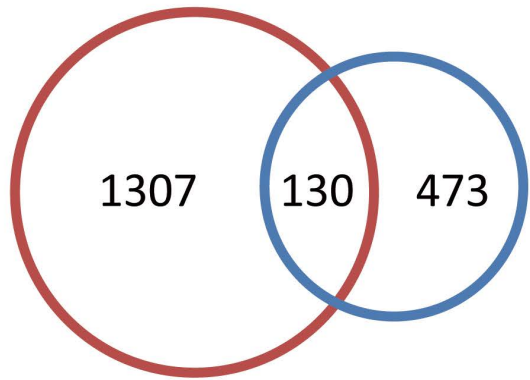
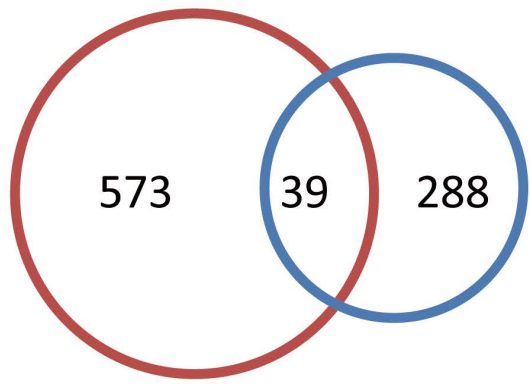
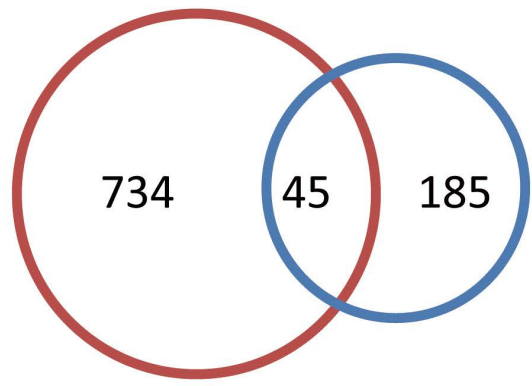

A
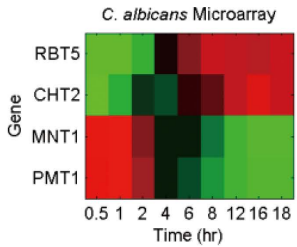
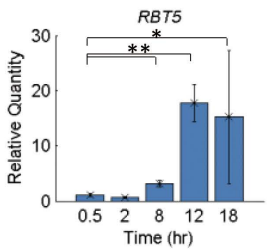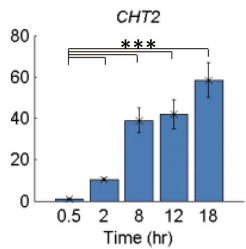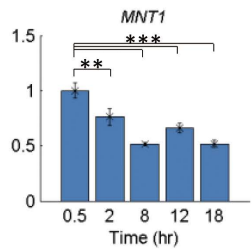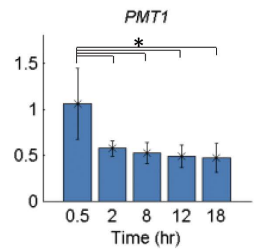

B
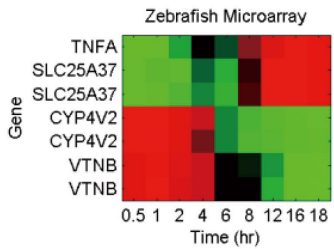
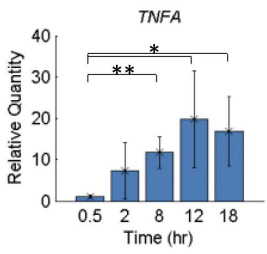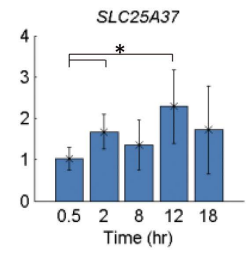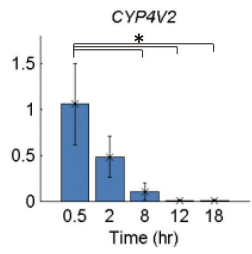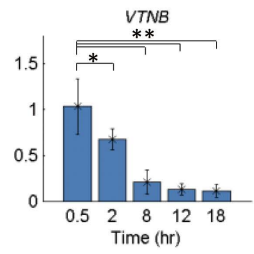

A 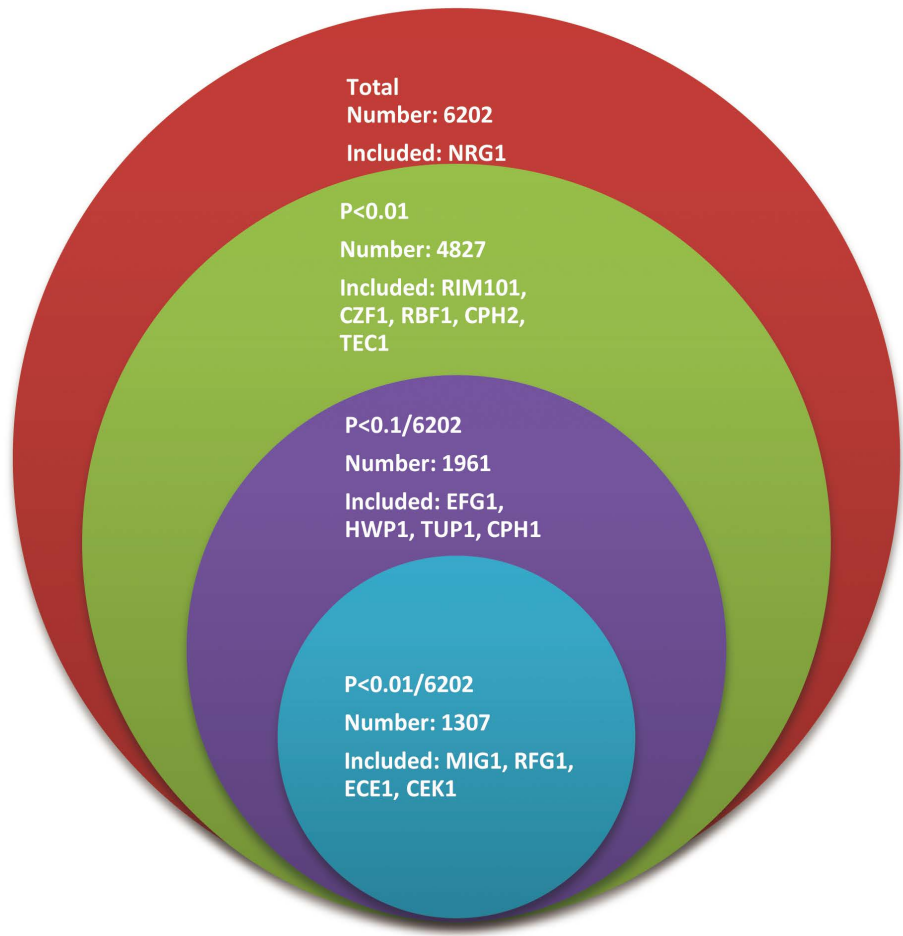

B 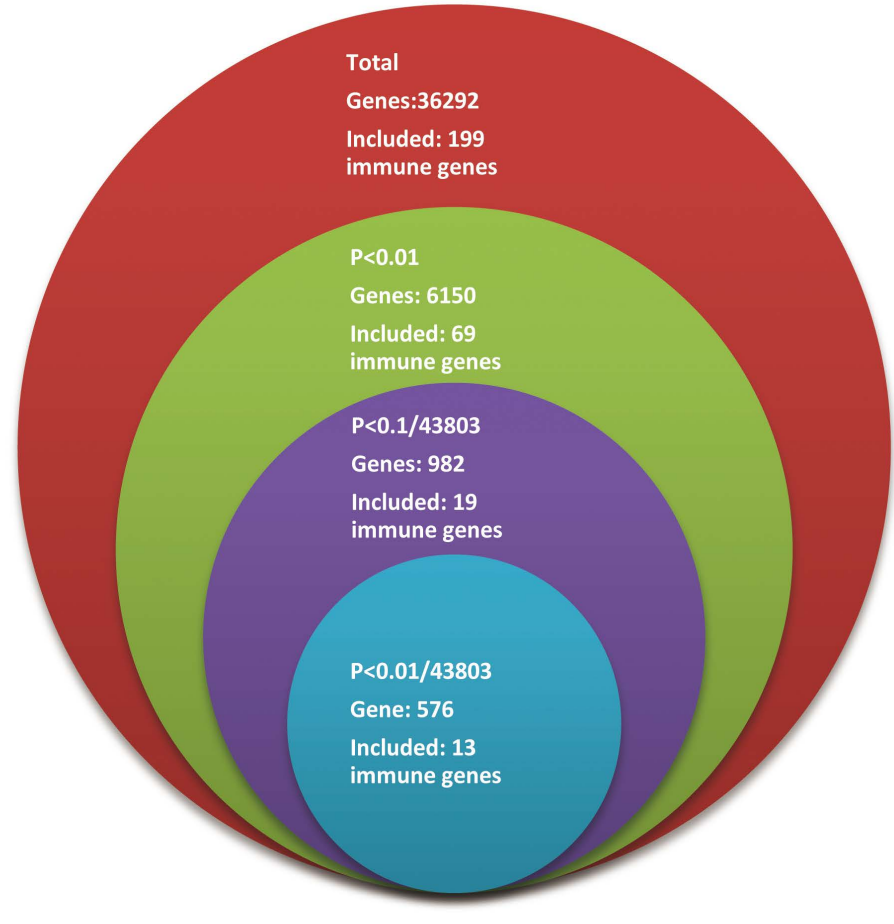

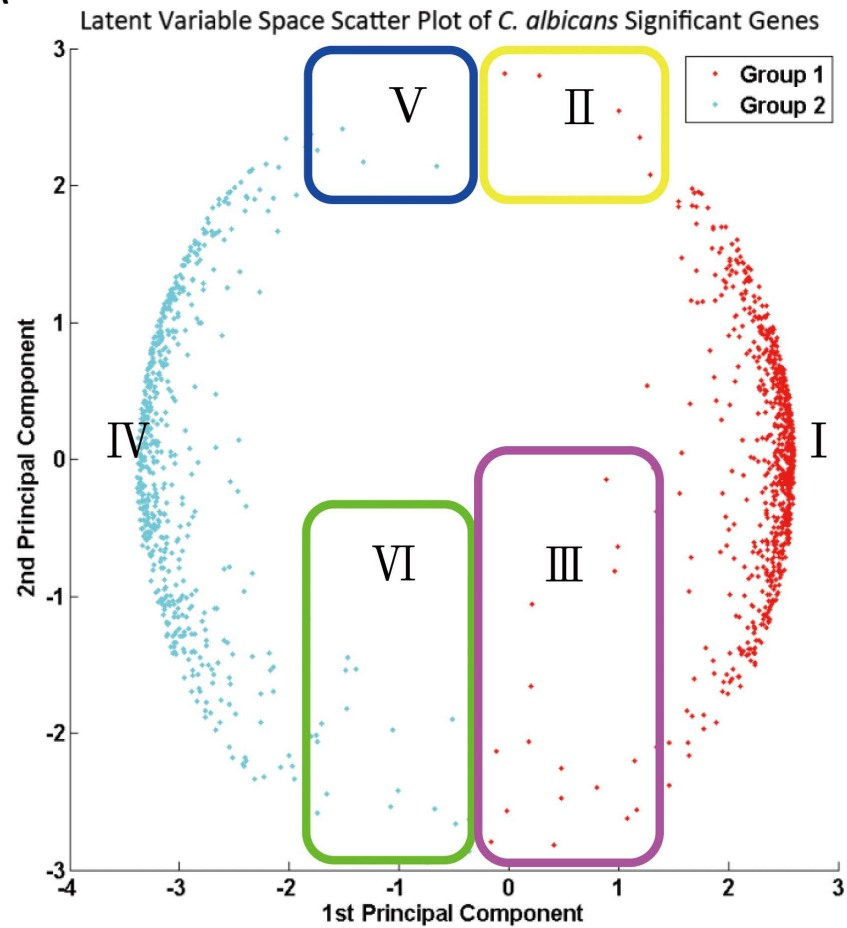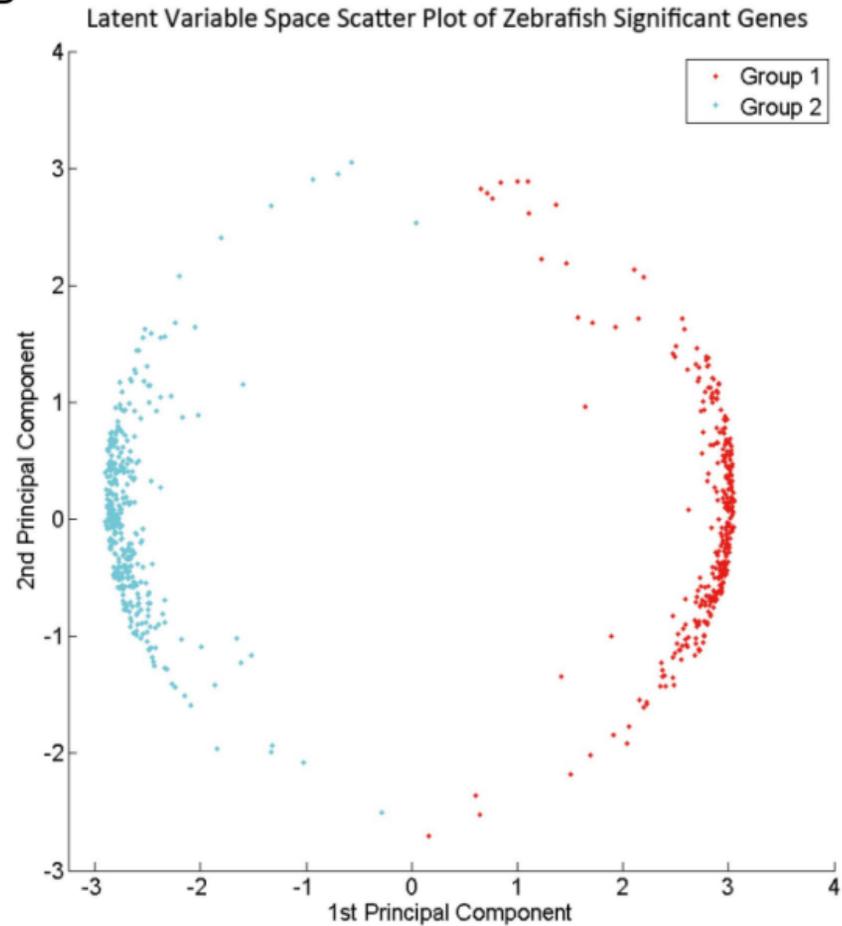